\documentstyle[preprint,aps]{revtex}

\begin{document}

\tighten
\draft

\title{Probing the Origin of the EMC Effect via Tagged Structure Functions
       of the Deuteron}

\author{W. Melnitchouk}
\address{Department of Physics,
        University of Maryland,
        College Park, MD 20742, USA}
\author{M. Sargsian}
\address{School of Physics and Astronomy, 
        Tel Aviv University, 
        Tel Aviv 69978, Israel, \\ 
        and 
        Yerevan Physics Institute,
        Yerevan 375036, Armenia}
\author{M.I. Strikman}
\address{Department of Physics, 
        Pennsylvania State University, 
        University Park, PA 16802, USA, \\
        and 
        Institute for Nuclear Physics, 
        St. Petersburg, Russia}

\maketitle

\begin{abstract}
We demonstrate that measurement of tagged structure functions of the 
deuteron in $(e,e'N)$ semi-inclusive reactions can discriminate between 
different hypotheses on the origin of the nuclear EMC effect.
By choosing extreme backward kinematics for the spectator nucleon to 
minimize effects from the deuteron wave function and final state 
interactions, one can isolate the modifications in the structure of 
the bound nucleon within the impulse approximation.
The same reaction can be used to extract the large-$x$ neutron to 
proton structure function ratio.
\end{abstract}

\section{Introduction}

More than a decade after the discovery of the nuclear EMC effect 
\cite{EMC1} and many fine measurements \cite{EMC2,EMC3,EMC4,EMC5} 
of the ratios of structure functions of nuclei and the deuteron, 
no consensus has been reached on the origin of the effect.
The $x$ dependence of the effect, while non-trivial, is rather smooth 
and has the same basic shape for all nuclei, making it is easy to fit
in a wide range of models with very different underlying assumptions.
The only extra constraint available so far comes for measurements of 
the $A$-dependence of the sea distribution, which restricts some of 
the models, but is still not sufficient to allow one to unambiguously
identify the origin of the EMC effect.

In order to move beyond this rather unsatisfactory situation, new 
experiments involving more kinematical variables accessible to 
accurate measurements are necessary.
The aim of this study is to demonstrate that use of semi-inclusive 
processes off the deuteron, 
\begin{equation}
\label{eD}
e + D \rightarrow e + N + X, 
\end{equation}
where a nucleon is detected in the target deuteron fragmentation 
region, may help to discriminate between some classes of models.
In particular, one may be able to distinguish between models in 
which the effect arises entirely from hadronic degrees of freedom 
--- nucleons and pions (which in the traditional nuclear physics
picture are responsible for the binding of the nucleus), 
and models in which the effect is attributed to the explicit 
deformation of the wave function of the bound nucleon itself.
By selecting recoil nucleons with small transverse momentum in 
the backward region, effects due to final state interactions (FSI)
can be minimized, thus allowing one to probe, within the impulse 
approximation, the deformation of the structure of the bound 
nucleon.

It is worth emphasizing that measurement of the reaction (\ref{eD})
in the kinematics of interest will become feasible in the near future,
for example at Jefferson Lab \cite{KG}.
Recent development of silicon detectors also make it possible in the 
jet target experiments to measure recoil nucleons with a low momentum 
threshold of about 100--150 MeV/c. 
Plans to implement such a technique are under discussion at the
HERMES detector at HERA \cite{HERA}.

Aside from providing insight into the origin of the nuclear EMC 
effect per se, the measurements of tagged events may also be useful 
in connection with the problem of extracting the neutron structure 
function from deuteron data.
By selecting only the slowest recoil protons in the target fragmentation
region, one should be able to isolate the situation whereby the virtual
photon scatters from a nearly on-shell neutron in the deuteron.
In this way one may hope to extract the $F_{2n}$ structure function
while minimizing uncertainties arising from modeling the nuclear effects 
in the deuteron.

This paper is organized as follows.
In Section II we outline the basic formalism for semi-inclusive 
deep-inelastic scattering off the deuteron, and discuss the conditions
under which the impulse approximation may be valid.
Section III is devoted to a survey of several models whose predictions
for the ratios of tagged structure functions are compared.
Possibilities of extracting the neutron structure function from 
semi-inclusive experiments are discussed in Section IV, and finally
some conclusions are drawn in Section V.

\section{Basic Formalism}

It was observed a long time ago that the nuclear EMC effect for the 
deviation from unity of the ratio
\begin{equation}
\label{remc}
R(x,Q^2) 
= {2 F_{2A}(x,Q^2)\over A F_{2D}(x,Q^2)} 
\end{equation}
is approximately proportional to the nuclear density.
This is natural for a dilute system, and indicates that most 
of the EMC effect is due to two-nucleon interactions.
Based on this observation one should expect that the EMC effect 
in the deuteron is much smaller than that in heavy nuclei. 
However, by virtue of the uncertainty principle, one may try to 
enhance the effect by isolating the configuration where the two 
nucleons in the deuteron are close together.
For example, it is easy to check that the main contributions to the 
deuteron wave function for nucleon momenta ${\bf p} \agt 300$ MeV/c
come from distances $\alt 1.2$~fm.

To make our discussion quantitative, let us begin by writing the 
electromagnetic tensor $W_D^{\mu\nu}$ of the deuteron in terms of 
the matrix element of the electromagnetic current: 
\begin{eqnarray}
W_D^{\mu\nu}
= \sum\limits_{spin, X} 
  \langle D\mid J^\mu_D(q)\mid X, N\rangle
  \langle X, N\mid J^{\mu\dag}_D(0)\mid D\rangle.
\end{eqnarray}
The tensor $W_D^{\mu\nu}$ can be expanded in terms of four possible
Lorentz structures, with the coefficients given by the invariant 
structure functions $F_L^D, F_T^D, F_{TL}^D$ and $F_{TT}^D$ 
($T=$ transverse, $L=$ longitudinal).
These functions can in general depend on four variables, constructed 
from the four-momenta of the target deuteron ($P$), virtual photon ($q$) 
and spectator nucleon ($p_s$) (or equivalently the momentum of the struck
nucleon $p$, where ${\bf p} = - {\bf p}_s$ in the deuteron rest frame).
In terms of the invariant structure functions the differential cross 
section for the semi-inclusive reaction $(e,e'N)$ can then be written:
\begin{eqnarray}
& & {d\sigma \over dx dQ^2 d^3{\bf p}_s/E_s} 
= {2\alpha_{em}^2\over x Q^4}
\left(1 - y - {x^2 y^2 M^2\over Q^2}\right)     
\nonumber \\ 
& & \times 
\left[ F_L^D
     + \left({Q^2\over 2{\bf q}^2} + \tan^2(\theta/2) \right)
        {\nu\over M} F_T^D
     + \left({Q^2\over {\bf q}^2} + \tan^2(\theta/2) \right)^{1\over 2} 
       \cos \phi\ F_{TL}^D
     + \cos (2\phi)\ F_{TT}^D
\right],
\label{e3}
\end{eqnarray}
where the structure functions are related to the components of the 
electromagnetic tensor $W_D^{\mu\nu}$ by:
\begin{mathletters}
\label{e4}
\begin{eqnarray}
\label{FLgood}
F_L^D    &=& \nu(1+\cos\delta)^2\cdot W_D^{- -},          \\
F_{T}^D  &=&  M (W_D^{xx} + W_D^{yy}),                    \\
F_{TL}^D &=&  2\nu (1+\cos\delta)\cdot W_D^{- x},         \\
F_{TT}^D &=&  {\nu\over 2} \sin^2\delta\cdot(W_D^{xx} - W_D^{yy}).
\end{eqnarray}
\end{mathletters}%
The kinematic variables in Eqs.(\ref{e3}) and (\ref{e4}) are 
$x = Q^2 / 2 M\nu$, where $\nu$ is the energy of virtual photon 
in the target rest frame and $M$ the nucleon mass,
$Q^2 = 4 E (E-\nu) \sin^2 (\theta/2)$ is the squared four-momentum 
transfer to the target, $E$ is the beam energy and $y = \nu/E$.
The angle $\phi$ is the azimuthal angle for the spectator nucleon,
and $E_s = \sqrt{M^2 + {\bf p}_s^2}$ the spectator nucleon energy, 
and $\sin^2\delta = Q^2 / {\bf q}^2$ in Eqs.(\ref{e4}).
We define the photon three-momentum ${\bf q}$ to be in the $+ z$ direction.

A convenient choice of the four independent variables for the structure 
functions is the two inclusive deep-inelastic scattering variables, $x$ 
and $Q^2$, and the transverse momentum, ${\bf p}_T^s$, and light-cone 
momentum fraction, $\alpha_s$, of the spectator nucleon:
\begin{eqnarray}
\alpha_s &=& {E_s-p_z^s \over M}, 
\end{eqnarray}
where $p^s_z$ is the longitudinal momentum of the detected nucleon 
(to simplify the expressions we have neglected here the deuteron 
binding energy, $M_D \approx 2 M$).
In the Bjorken limit the variable $\alpha_s$ then satisfies the 
condition \cite{FS81}:
\begin{equation}
\label{alp}
\alpha_s \le 2 - x. 
\end{equation}

Having defined the relevant cross sections, we must now establish the
kinematical range in which the nuclear modifications of the bound nucleon
structure function will be accessible experimentally.
{}From Eqs.(\ref{e3}) and (\ref{e4}) the restrictions:
\begin{mathletters}
\label{e5}
\begin{eqnarray}
& & Q^2/{\bf q}^2 \sim 4 M^2 x/Q^2\ll 1,          \\ 
\label{e5a}
& & Q^2 \ll E(E-\nu), 
\label{e5b}
\end{eqnarray}
\end{mathletters}%
enhance the contribution of the longitudinal structure function $F_L$, 
which is expressed through the ``good'' component of the electromagnetic
current, Eq.(\ref{FLgood}).
Another simplification can be achieved by considering the situation 
where the detected nucleon in the deuteron is in the spectator 
kinematics, namely: 
\begin{equation}
\alpha_s \ge 1-x.
\label{skin}
\end{equation}
In this kinematical region the contribution of the direct process 
where a nucleon is produced at the $\gamma^* N$ interaction vertex
is negligible \cite{FS81}.

\subsection{Factorization and the Impulse Approximation}
\label{FIA} 

In formulating deep-inelastic scattering from the deuteron the
simplest approach adopted has been the impulse approximation
for the nuclear system. 
We will consider two formulations of the impulse approximation: 
one based on the covariant Feynman, or instant-form, approach, 
where one nucleon is on-mass-shell and one off-mass-shell, and 
the non-covariant/Hamiltonian light-cone approach, in which both 
nucleons are on-mass-shell (but off the light-cone energy shell).
In the next subsection we shall consider corrections to the impulse 
approximation, in the form of final state interactions, but for now 
let us review briefly the basic impulse approximation assumptions 
and results.

\subsubsection{Covariant Instant-Form Approach}
\label{IF}

In order to construct covariant amplitudes in the instant-form of 
quantization, summation over all possible time-orderings of 
intermediate states is essential. 
Incorporation of negative energy configurations into the total 
Lorentz-invariant amplitude is done on the basis of introducing
intermediate state particles which are off their mass shells.
In the impulse approximation for the scattering of a virtual photon, 
$\gamma^*$, from the deuteron, the invariant amplitude for the 
complete process factorizes into a product of amplitudes for 
$\gamma^*$--off-mass-shell nucleon scattering, and for forward 
nucleon--deuteron scattering.
In this case the hadronic tensor of the deuteron can be written 
\cite{MST}:
\begin{eqnarray}
\label{DIA} 
W^{\mu\nu}_D(P,p,q)
&=& {\rm Tr} 
\left[ {\cal\widehat S}_{ND}(P,p)\ 
        \widehat W^{\mu\nu}_N(p,q)
\right], 
\end{eqnarray}
where the nucleon--deuteron scattering amplitude ${\cal\widehat S}_{ND}$
(for spin-averaged processes) contains scalar and vector components:
\begin{eqnarray}
{\cal\widehat S}_{ND}
&=& {\cal\widehat S}_0\ 
 +\ \gamma_{\alpha} {\cal\widehat S}^{\alpha}_1. 
\end{eqnarray}
The operator $\widehat W^{\mu\nu}_N$ in Eq.(\ref{DIA}) is the truncated 
hadronic tensor for the off-shell nucleon, which describes the $\gamma^* N$
interaction.
Because $\widehat W^{\mu\nu}_N$ is a matrix in both Lorentz and Dirac spaces,
its structure is necessarily more general than that for a free nucleon.
In the Bjorken limit, it was shown in Ref.\cite{MST} that out of a possible
13, only three independent structures contribute:
\begin{eqnarray}
\label{Woff}
\widehat W_N^{\mu\nu}(p,q)
&=& - \left( g_{\mu\nu} + {q_{\mu} q_{\nu} \over Q^2} \right)
\left(  \widehat W_0(p,q)
     +  \not\!p\ \widehat W_1(p,q)
     +  \not\!q\ \widehat W_2(p,q)
\right)\ +\ {\cal O}\left( {1 \over Q^2} \right).
\end{eqnarray}
Note that the full expression for $\widehat W_N^{\mu\nu}$ contains
both positive and negative energy pieces.
The on-shell nucleon tensor is obtained from $\widehat W_N^{\mu\nu}$
by projecting out the positive energy components:
\begin{mathletters}
\begin{eqnarray}
W^{\mu\nu}_N(p,q)
&=& {1\over 4} {\rm Tr}
\left[ \left(\not\!p + M \right) \widehat W_{\mu\nu}(p,q)
\right]                                                         \\
&=& - \left( g_{\mu\nu} + {q_{\mu} q_{\nu} \over Q^2} \right)
\left( M \widehat W_0
     + M^2 \widehat W_1
     + p\cdot q \widehat W_2
\right), 
\end{eqnarray}
\end{mathletters}%
where the functions $\widehat W_{0\cdots 2}$ here are evaluated 
at their on-shell points.

In terms of the off-shell truncated functions $\widehat W_{0\cdots 2}$
the deuteron tensor can then be written:
\begin{eqnarray}
W^{\mu\nu}_D(P,p,q)
&=& -4 \left( g_{\mu\nu} + {q_{\mu} q_{\nu} \over Q^2} \right)
\left( {\cal\widehat S}_0\ \widehat W_0\
    +\ {\cal\widehat S}\cdot p\ \widehat W_1\ 
    +\ {\cal\widehat S}\cdot q\ \widehat W_2
\right), 
\end{eqnarray}
so that in general the right-hand-side does not factorize into 
a single term with separate nuclear and nucleon components. 
As explained in Ref.\cite{MST}, factorization can be recovered by 
projecting from ${\cal\widehat S}_{0,1}$ only the positive energy
(on-shell) components, in which case: 
\begin{equation}
{\cal\widehat S}_0\ \propto
{\cal\widehat S}\cdot p\ \propto
{\cal\widehat S}\cdot q. 
\end{equation}
This proportionality can also be obtained by taking the non-relativistic
limit for the $ND$ amplitudes to order ${\bf p}^2/M^2$ \cite{KPW}, 
since all negative energy contributions enter at higher orders 
\cite{MSTD,KMPW}.
For the relativistic $N D$ interaction, the negative energy components 
enter through the $P$-state wave functions of the deuteron.
However, in most realistic calculations of these, in the context of
relativistic meson-exchange models of the deuteron \cite{RELWFN}, 
the contribution of the $P$-state wave functions is only a fraction of
a percent, and so long as one avoids regions of extreme kinematics
which are sensitive to the very large momentum components of the 
deuteron wave function, dropping the negative energy components is
a reasonable approximation.
Indeed, the explicit calculations of the inclusive structure function
of the deuteron with relativistic wave functions shows that the 
factorization-breaking corrections amount to $\alt$ 1--2\% for all values 
of $x \alt 0.9$ \cite{MSTD,MPT}.

Finally, given these approximations one can write the deuteron hadronic
tensor as:
\begin{eqnarray}
\label{ia}
W^{\mu\nu}_D(P,p,q)
&\approx& S^{IF}(P,p)\ W^{\mu\nu\ eff}_N(p,q),
\end{eqnarray}
where $S^{IF}$ is the nucleon spectral function within the instant-form 
impulse approximation \cite{MSTD}:
\begin{eqnarray}
\label{Sif}
S^{IF}(\alpha_s,p_T)
&=& (2-\alpha_s) {E_s M_D \over 2(M_D - E_s)}
| \psi_D(\alpha_s,p_T) |^2,
\end{eqnarray}
normalized such that
\begin{eqnarray}
\int d^2{\bf p}_T {d\alpha_s \over \alpha_s} S^{IF}(\alpha_s,p_T)
&=& 1
\end{eqnarray}
to ensure baryon number conservation \cite{FS76,LP78}.
Note that the approximation for the spectral function in Eq.(\ref{Sif})
is valid only for non-relativistic momenta, and for large momenta the
full (non-convolution) expression for the hadronic tensor in Ref.\cite{MSTD}
should be used.
The effective nucleon hadronic tensor $W^{\mu\nu\ eff}_N$ is defined as:
\begin{eqnarray}
W^{\mu\nu\ eff}_N
&=& - \left( g_{\mu\nu} + {q_{\mu} q_{\nu} \over Q^2} \right)
\left( M \widehat W_0\
    +\ p^2 \widehat W_1\ 
    +\ p \cdot q \widehat W_2
\right) 
\nonumber\\
&\equiv& -\left( g_{\mu\nu} + {q_{\mu} q_{\nu} \over Q^2} \right)
F_{1N}^{eff}\left( {x \over 2-\alpha_s},p^2,Q^2 \right)\ + \cdots ,
\end{eqnarray}
where the effective nucleon structure function $F_{1N}^{eff}$ is now
a function of the momentum fraction $x/(2-\alpha_s)$ and the virtuality
$p^2$ of the bound nucleon, as well as $Q^2$.
The kinematics of the spectator process gives rise to the relation: 
\begin{eqnarray}
p^2 &=& -{2 p_T^2 + (2-\alpha_s) M^2 \over \alpha_s}\ 
     +\ {1 \over 2} (2-\alpha_s) M_D^2\ , 
\label{virt}
\end{eqnarray}
using which one can equivalently express $F_{1N}^{eff}$ as a function 
of the transverse momentum $p_T$ of the interacting nucleon, rather 
than $p^2$.

\subsubsection{Light-Cone Approach}

In the light-cone formalism one can formally avoid the problems
associated with negative energy solutions in Eq.(\ref{Woff}), 
however, to obtain Eq.(\ref{ia}) one must on the other hand consider
contributions arising from instantaneous interactions \cite{FS88}. 
To take these effects into account one has to use gauge invariance 
to express the contribution of the ``bad'' current components of the 
electromagnetic tensor through the ``good'' components:
\begin{equation}
J^A_+ = - {q_+\over q_-}J^A_-, 
\end{equation}
and include the contribution of the instantaneous exchanges using 
the prescription of Brodsky and Lepage \cite{BROD}.
In the approximation when other than two-nucleon degrees of freedom 
in the deuteron wave function can be neglected, one can unambiguously
relate the light-cone deuteron wave functions to those calculated 
in the rest frame in non-relativistic instant-form calculations 
\cite{PARIS,BONN}.
The final result for the spectral function in the light-cone approach
can be written similarly to Eq.(\ref{ia}), only now $S$ is replaced 
by the light-cone density matrix:
\cite{FS81,FS88},
\begin{equation}
S^{LC}(\alpha_s,p_T) = { \sqrt{M^2 + {\bf k}^2} \over 2-\alpha_s }
                        |\psi_D(k)|^2, 
\label{lcs}
\end{equation}
where 
\begin{equation}
k \equiv |{\bf k}|
= \sqrt{{M^2 + p_T^2\over \alpha_s (2-\alpha_s)} - M^2}
\label{k}
\end{equation}
is the relative momentum of the two nucleons on the light-cone,
and $W^{\mu\nu\ eff}_N$ in Eq.(\ref{ia}) is now the bound nucleon 
electromagnetic tensor defined on the light-cone.
Note that the main operational difference between the instant-form 
(\ref{Sif}) and light-cone (\ref{lcs}) impulse approximations is the 
different relation between the deuteron wave function and the scattering
amplitude.
Numerical studies have demonstrated that in the kinematical region of 
interest, $|{\bf p}_s| \alt 0.5$ GeV/c, the difference between the 
results of the two approximations (for the same deuteron wave function)
is quite small --- see Sec.\ref{numcom} and Ref.\cite{FS81}.

The structure functions for the scattering from the off-``+''-shell 
bound nucleons may depend on the variables of this nucleon similarly 
to the case of the Bethe-Salpeter, or covariant Feynman, approach. 
In another language this dependence can be interpreted as the presence
of non-nucleonic degrees of freedom in the deuteron.
With this is mind, we shall use Eq.(\ref{ia}) as the basis for the 
results discussed in the following sections.
Before focusing on specific model calculations of the semi-inclusive
deuteron structure functions, however, let us first turn our attention 
to the validity of the impulse approximation, and the problem of final 
state interactions in particular.

\subsection{Final State Interactions}

In the kinematical region defined by Eq.(\ref{skin}) the contribution
of direct processes, where a nucleon is produced in the $\gamma^* N$ 
interaction, is negligible \cite{FS81}. 
Therefore within the framework of the distorted wave impulse 
approximation (DWIA)
\footnote{Note that the DWIA approach works best at extreme backward 
kinematics, Eq.(\ref{pt}) below, where final state interactions have
a small contribution, Eq.(\ref{e7}).}
the total deuteron tensor $W^{\mu\nu}_D$ can be expressed through 
the nucleon electromagnetic currents as:
\begin{eqnarray}
W^{\mu\nu}_D(x,\alpha_s,p_T,Q^2) 
&\approx& 
        \left| \sum \langle D|p n \rangle 
        \langle X N \mid {\cal O}_{IA} + {\cal O}_{FSI} \mid X N 
        \rangle\right|^2\ W^{\mu\nu}_N(x,\alpha_s,p_T,Q^2)
\nonumber\\
&\equiv& S^{DWIA}(\alpha_s,p_T)\ W^{\mu\nu\ eff}_N(x,\alpha_s,p_T,Q^2), 
\label{e6}
\end{eqnarray}
where ${\cal O}_{IA}$ is the impulse approximation operator, while 
${\cal O}_{FSI}$ describes the soft final state interactions between 
the final hadronic products and the spectator nucleon. 
The function $S^{DWIA}(\alpha_s,p_T)$ now represents the spectral 
function distorted by FSI effects.

Analysis \cite{STZ} of the recent high energy deep-inelastic scattering 
data on slow neutron production \cite{E665} is rather indicative that 
even in heavy nuclei final state interactions are small, so that the 
average number of hadrons which reinteract in the target does not 
exceed unity.
This indicates that the system which is produced in the $\gamma^* N$ 
interaction is quite coherent and interacts at high energies with 
a relatively small effective cross section, 
\begin{eqnarray}
\sigma_{eff} &\ll& \sigma_{NN}. 
\end{eqnarray}
Such a situation allows one to factorize the $\gamma^* N$ interaction
and use the calculation of FSI in $e D \rightarrow e\ p\ n$ processes
as a conservative upper limit.

The final simplification which we can gain is to consider the extreme 
backward kinematics where, in addition to Eq.(\ref{skin}), we also require:
\begin{equation}
{\bf p}_T \approx 0. 
\label{pt}
\end{equation}
In this case it can be shown \cite{FMGSS95,FMGSS96} that:
\begin{equation}
S^{DWIA}(\alpha_s,p_T \approx 0) 
\sim S(\alpha_s,p_T \approx 0)
\left[ 1 - {\sigma_{eff}(Q^2,x) \over 8\pi <r_{pn}^2>}
           {|\psi_D(\alpha_s,<p_T>) \psi_D(\alpha_s,0)| \over
             S(\alpha_s,p_T \approx 0) / \sqrt{E_s\ E_s(<p^2_T>)}}
\right], 
\label{e7}
\end{equation}
where $<r_{pn}^2>$ is the average separation of the nucleons within the
deuteron, $E_s$ is the spectator nucleon energy, and
$E_s(<p^2_T>)=\sqrt{M^2 + p_z^{s\ 2} + <p^2_T>}$ is the energy evaluated
at the average transverse momentum $<p^2_T>^{1/2} \sim$ 200--300 MeV/c 
transferred for the hadronic soft interactions with effective cross 
section $\sigma_{eff}$. 
The steep momentum dependence of the deuteron wave function, 
$|\psi_D(\alpha_s,<p_T>)| \ll |\psi_D(\alpha_s,p_T \approx 0)|$,
ensures that FSI effects are suppressed in the extreme backward kinematics,
in which case the original impulse approximation expression for 
$S(\alpha_s,p_T)$ can be used rather than $S^{DWIA}(\alpha_s,p_T)$.
Finally, expressing the electromagnetic tensor of the nucleon, 
$W^{\mu\nu}_N$, through the effective nucleon structure function
$F_{1N}^{eff}$ and $F_{2N}^{eff}$, we can then write for the 
deuteron tensor:
\begin{eqnarray}
W_D^{\mu\nu} 
&\approx& S(\alpha_s,p_T) 
\left\{ 
- \left( g_{\mu\nu} + {1 \over Q^2} q_{\mu} q_{\nu} \right)
 {1 \over M} F_{1N}^{eff}\left({x \over 2-\alpha_s},p_T,Q^2\right)
\right.
\nonumber\\
& & \hspace*{1cm}
\left.
 + \left( p_{\mu}+{p \cdot q\over Q^2} q_{\mu} \right)
   \left( p_{\nu}+{p \cdot q\over Q^2} q_{\nu} \right)
   {1 \over \nu M^2}
   F_{2N}^{eff}\left({x \over 2-\alpha_s},p_T,Q^2\right) 
\right\}.
\label{tn3}
\end{eqnarray}
Defining $F_{L,T,TL,TT}^N$ to be the semi-inclusive structure 
functions in Eqs.(\ref{e4}) with the spectral function factored, 
\begin{eqnarray}
F_{L,T,TL,TT}^D
&=& S(\alpha_s,p_T)\ F_{L,T,TL,TT}^N\ ,
\end{eqnarray}
one can express the semi-inclusive nucleon functions in terms 
of the effective structure functions of the nucleon as: 
\begin{mathletters}
\label{e8}
\begin{eqnarray}
F_L^N(x,\alpha_s,p_T,Q^2) 
&=& -\sin^2\delta {\nu\over M}\ 
        F_{1N}^{eff}\left({x \over 2-\alpha_s},p_T,Q^2\right)   \nonumber\\
&+& (1+\cos\delta)^2 
    \left(\alpha_s + {p \cdot q\over Q^2} \alpha_q\right)^2
    {\nu \over \widetilde \nu} 
        F_{2N}^{eff}\left({x \over 2-\alpha_s},p_T,Q^2\right),          \\
F_T^N(x,\alpha_s,p_T,Q^2) 
&=& 2 F_{1N}^{eff}\left({x \over 2-\alpha_s},p_T,Q^2\right)
 + {p_T^2 \over M^2} {M \over \widetilde \nu} 
        F_{2N}^{eff}\left({x \over 2-\alpha_s},p_T,Q^2\right),          \\ 
F_{TL}^N(x,\alpha_s,p_T,Q^2) 
&=& 2 (1+\cos\delta) {p_T \over M}
    \left( \alpha_s + {p \cdot q\over Q^2} \alpha_q \right)
    {\nu \over \widetilde \nu} 
        F_{2N}^{eff}\left({x \over 2-\alpha_s},p_T,Q^2\right),          \\
F_{TT}^N(x,\alpha_s,p_T,Q^2) 
&=& {\sin^2\delta \over 2}
    {p_T^2 \over M^2}{\nu \over \widetilde \nu} 
        F_{2N}^{eff}\left({x \over 2-\alpha_s},p_T,Q^2\right), 
\end{eqnarray}
\end{mathletters}%
where 
\begin{mathletters}
\label{tnu}
\begin{eqnarray}
\alpha_q &\equiv& {\nu - |{\bf q}|\over M},             \\
\widetilde \nu &\equiv& {p \cdot q\over M}\ 
=\ |{\bf q}|{1+\cos\delta\over 2}\alpha_s 
 + \alpha_q {M^2+p_T^2 \over 2\alpha_s M}.
\end{eqnarray}
\end{mathletters}%
Note that within impulse approximation the Callan-Gross relation 
between the $F_{1N}^{eff}$ and $F_{2N}^{eff}$ structure functions  
is preserved.
Therefore the experimental verification of this relation could serve 
as another way to identify FSI effects.

Equations (\ref{e8}) and (\ref{tnu}) show that at fixed $x$ and 
$Q^2 \rightarrow \infty$, when $\alpha_q\rightarrow 0$ the longitudinal
structure function, $F_L^N$, does not depend explicitly on the transverse 
momentum of the nucleon. 
It contains $p_T$ dependence only in the argument of the bound nucleon 
structure functions, which arises from the possible nuclear modifications 
of the nucleon's parton distributions.
The above argument, and the fact that FSI of hadronic products with 
$p_X \agt 1$ GeV/c practically conserve $\alpha_s$, allows one to conclude
that the FSI effect on $F_L^N$ is minimal.

On the other hand the functions $F_{T}$, $F_{TL}$ and $F_{TT}$ do 
explicitly depend on the spectator transverse momentum and therefore 
the hadronic reinteractions in the final state may strongly affect
these structure functions. 
Such a situation suggests that a separate study of the complete set 
of the structure functions will allow one to investigate the effects 
of final state interactions in the deep-inelastic $(e,e'N)$ reactions.  
Note that for the production of spectators with $p_T > 0$, FSI effects
are not likely to depend strongly on $x$ for $x>0.1$.
This is because at $x > 0.1$ the essential longitudinal distances in
deep-inelastic scattering are small. 
Therefore for these kinematics one expects that
$\sigma_{eff}(Q^2,x) \approx \sigma_{eff}(Q^2)$.
The effect of FSI will be investigated in the dedicated deep-inelastic
scattering experiments at HERMES which will measure the $A$-dependence
of the forward produced hadrons.

These observations enable us to conclude that in the kinematic region 
defined by Eqs.(\ref{e5}), (\ref{skin}) and (\ref{pt}), where the 
contribution of $F_L^N$ is enhanced and FSI effects are small, 
the differential cross section (\ref{e3}) can be written:
\begin{eqnarray}
{d\sigma^{e D \rightarrow e p X} \over 
 dx dW^2 d(\log\alpha_s) d^2{\bf p}_T } 
&\approx& {2\alpha_{em}^2 \over Q^4} (1-y) {S(\alpha_s,p_T)\over 2-\alpha_s}
\left[ F_L^N + {Q^2\over 2{\bf q}^2}{\nu\over M} F_T^N \right] 
\nonumber \\ 
&=& {2\alpha_{em}^2 \over Q^4} (1-y) S(\alpha_s,p_T) 
    F^{eff}_{2N}\left({x\over 2-\alpha_s},p_T,Q^2\right)
\label{e9}
\end{eqnarray}
where we have made the transformation 
$dQ^2/x \rightarrow dW^2$, with $W^2=-Q^2+2M\nu+M^2$.

Based on the expectation that FSI effects should not strongly depend on $x$,
from Eq.(\ref{e9}) it may be advantageous to consider the ratio of cross
sections relative to a given $x$, in the range 0.1--0.2, where the observed
EMC effect in inclusive scattering is small \cite{FS88,FS85}:
\begin{eqnarray} 
\label{G}
G(\alpha_s,p_T,x_1,x_2,Q^2)
&\equiv& 
\left.
{ d\sigma (x_1,\alpha_s,p_T,Q^2) \over 
  dx dW^2 d(\log\alpha_s) d^2{\bf p}_T } 
\right/
{ d\sigma (x_2,\alpha_s,p_T,Q^2) \over 
  dx dW^2 d(\log\alpha_s) d^2{\bf p}_T } 
\nonumber\\
&=& {F^{eff}_{2N}(x_1/(2-\alpha_s),p_T,Q^2) \over 
     F^{eff}_{2N}(x_2/(2-\alpha_s),p_T,Q^2)}. 
\end{eqnarray}
In our analysis we will consider only production of backward nucleons, 
$\alpha_s \ge 1$, to suppress contributions from the direct processes 
where a nucleon is produced in the $\gamma^* N$ interaction vertex.
The more liberal condition, Eq.(\ref{skin}), is in reality sufficient
\cite{FS81}.

Note also that for heavier nuclei the FSI becomes much more important
in the limit of large $x$. 
As was demonstrated in Ref.\cite{FS88}, in this limit rescattering of 
hadrons produced in the elementary deep-inelastic scattering off the 
short-range correlation is dynamically enhanced, since the average value 
of the Bjorken-variable for this mechanism is $\approx x$, as opposed 
to $x / (2-\alpha_s)$ for the spectator mechanism.
In this sense the deuteron target provides the best way of looking for 
the EMC effect for bound nucleons.

\section{Models}

In this section we briefly summarize several models of the EMC effect
which we use in our analysis and present their predictions for the tagged
structure functions.
The differences between the models stem from dynamical assumptions about
the deformation of the bound nucleon wave functions, and from the fraction
of the EMC effect attributed to non-baryonic (mesonic) degrees of freedom
in nuclei --- from the dominant part in some versions of the binding
model, to the models where non-baryonic degrees of freedom play no role,
as in the color screening or QCD radiation models.
(Other models which have been used in studies of semi-inclusive DIS include
the six-quark cluster models discussed in Refs.\cite{CLS,CS}.)

\subsection{Binding Models}

One of the simplest of the early ideas proposed to explain the nuclear EMC 
effect was the nuclear binding model, in which the main features of the EMC
effect could be understood in terms of conventional nuclear degrees of 
freedom --- nucleons and pions --- responsible for the binding in nuclei
\cite{FS88,AKV,DT,ET,ANL,HM,CL,MEC,BT}.
Within the formalism of Sec.\ref{IF}, the inclusive nuclear structure 
function in the EMC ratio, Eq.(\ref{remc}), is expressed through a 
convolution of the nuclear spectral function and the structure function 
of the bound (off-shell) nucleon (c.f. Eq.(\ref{ia})).
Contributions from DIS from the pionic fields themselves, which are 
needed to balance overall momentum conservation, were considered in 
Refs.\cite{AKV,ET,ANL,HM,MEC}, however, their role is most evident only
at small $x$ ($x \alt 0.2$).

The bulk of the suppression of the EMC ratio (\ref{remc}) at $x \sim 0.6$
in the binding model can be attributed to the fact that the average value 
for the interacting nucleon light-cone fraction is less than unity.
For the case of the deuteron, this corresponds to the average spectator
light-cone fraction $\langle\alpha_s\rangle > 1$, which is contrary to 
what one would have from Fermi motion alone, where the average $\alpha_s$ 
is $< 1$.
A relatively minor role is played by the structure function of the bound
nucleon itself --- the only requirement is that it be a monotonically
decreasing function of $x$ \cite{FS81,AKV,BT}.
This is clearly the case for the on-shell structure function, and since 
the off-shell behavior of the bound nucleon structure function is unknown, 
most early versions of the binding model simply neglected the possible 
dependence on $p^2$ with the expectation that it is not large for weakly 
bound systems.
Only very recently has the issue of off-shell dependence in the bound 
nucleon structure function been addressed \cite{MST,KPW}, where the first 
attempts to construct models for the $p^2$ dependence of $F_{2N}^{eff}$ 
were made.
Note that a consequence of assuming the absence of any off-shell effects
in $F_{2N}^{eff}$ is that the tagged structure function ratio $G$ in 
Eq.(\ref{G}), normalized to the corresponding ratio for a free proton, 
would be unity (see Fig.5 below).
Any observed deviation of this ratio from unity would therefore be a 
signal of the presence of nucleon off-shell effects.

If in a dilute system such as the deuteron the nucleon off-shell effects 
do not play a major role (at least at $x \alt 0.7$), one could expand the 
effective nucleon structure function in a Taylor series about $p^2 = M^2$:
\begin{eqnarray}
\label{p2expand}
F_{2N}^{eff}(x,p^2,Q^2) 
&=& F_{2N}(x,Q^2)\ 
 +\ (p^2-M^2) 
    \left. { \partial F_{2N}^{eff}(x,p^2,Q^2) \over \partial p^2 }
    \right|_{p^2=M^2}
 +\ \cdots .
\end{eqnarray}
Here the off-shell dependence is determined, to order ${\bf p}^2/M^2$,
entirely by the derivative of $F_{2N}^{eff}$ with respect to $p^2$.
In order to model this correction a microscopic model of the nucleon 
structure is required \cite{MST,KPW,KMPW}.
In any generic quark-parton model, the effective nucleon structure function 
can be written as an integral over the quark momentum $p_q$ of the quark 
spectral function $\rho$:
\begin{eqnarray}
\label{rho}
F_{2N}^{eff}(x,p^2,Q^2) &=& \int dp_q^2\ \rho(p_q^2,p^2,x,Q^2). 
\end{eqnarray}
To proceed from Eq.(\ref{rho}) requires additional assumptions about the 
quark spectral function.
The simplest is to assume that the $p^2$ and $p_q^2$ dependence in $\rho$ 
is factored \cite{KPW,KMPW}, which then leads to an explicit constraint on
the $p^2$ derivative of $F_{2N}^{eff}$ from baryon number conservation in 
the deuteron: 
\begin{eqnarray}
\int_0^1 {dx \over x} 
\left. { \partial F_{2N}^{eff}(x,p^2,Q^2) \over \partial p^2 }
\right|_{p^2=M^2} &=& 0.
\end{eqnarray}
This then allows $\partial F_{2N}^{eff}(x,p^2,Q^2) / \partial p^2$
to be determined from the $x$-dependence of the on-shell structure 
function $F_{2N}(x,Q^2)$ in terms of a single parameter, the squared
mass of the intermediate state ``diquark'' system that is spectator 
to the deep-inelastic collision, $(p-p_q)^2$.
One can obtain a good fit to the on-shell nucleon structure function 
data in terms of this model if one restricts the spectator ``diquark''
mass to be in the range $(p-p_q)^2 \approx 2-4$ GeV$^2$ \cite{KPW}.

A more microscopic model which does not rely on the factorization 
of the $p^2$ and $p_q^2$ dependence in $\rho$ was discussed in 
Ref.\cite{MST}.
The quark spectral function there was determined entirely from 
the dynamics contained in the nucleon--quark--spectator ``diquark''
vertex function, $\Gamma(p,p_q)$. 
Within the approximation discussed in Section \ref{FIA}, taking 
the positive energy nucleon projection only, 
\begin{eqnarray}
\rho(p_q^2,p^2,x,Q^2) &\rightarrow& 
{\rm Tr} 
\left[ (\not\!p + M)\ \overline\Gamma(p,p_q)\ 
       (\not\!p_q - m_q)^{-1}\ \not\!q\ (\not\!p_q - m_q)^{-1}\
       \Gamma(p,p_q)
\right], 
\end{eqnarray}
where $m_q$ is the quark mass. 
Angular momentum conservation allows two forms for the vertex function 
$\Gamma$, namely scalar and pseudo-vector.
In Refs.\cite{MST,MSTD,MPT} it was found that the on-shell data could
be well described in terms of only a few of the many possible Dirac 
structures for $\Gamma$.
In particular, the vertices were chosen to be $\propto I$ and 
$\gamma_{\alpha} \gamma_5$.
The momentum dependence of the vertex functions, on the other hand, is 
more difficult to derive, and must in practice be either parameterized 
or calculated by solving bound state Faddeev equations in simple models
of the nucleon \cite{FADVF}. 
In order to obtain realistic static properties of the nucleon, and to 
account for the bound nature of the nucleon state, the vertex functions
must have the form: 
\begin{eqnarray}
\Gamma(p,p_q) 
&\propto& { \left( m_q^2 - p_q^2 \right) 
      \over \left( \Lambda^2 - p_q^2 \right)^n },
\end{eqnarray}
where the parameters $\Lambda$ and $n$ are fixed by comparing with the 
quark distribution data, and the overall normalization is fixed by the 
baryon number conservation condition for both the nucleon and deuteron 
structure functions \cite{MST,MSTD,MPT}.
In the comparisons in Sec.\ref{numcom} we use the parameters from the 
analysis of Ref.\cite{MST}.

\subsection{Color Screening Model of Suppression of Point-like 
            Configurations in Bound Nucleons}

A significant EMC effect in inclusive $(e,e')$ reactions occurs for 
$x\sim 0.5$--0.6 which corresponds to the high-momentum component of 
the quark distribution in the nucleon.
Therefore the EMC effect in this $x$ range is sensitive to a rather 
rare component of the nucleon wave function where 3 quarks are likely
to be close together\cite{FS88,FS85}. 
It is assumed in this model that for large $x$ the dominant contribution
to $F_{2N}(x,Q^2)$ is given by the point-like configurations (PLC) of
partons which weakly interact with the other nucleons.
Note that due to scaling violation $F_{2N}(x,Q^2)$ at $x \gtrsim 0.6$, 
$Q^2 \gtrsim 10$~GeV$^2$, is determined by the nonperturbative nucleon 
wave function at $x \gtrsim 0.7$. 
Thus it is actually assumed that in the nonperturbative nucleon wave 
function point-like configurations dominate at $x \gtrsim 0.7$.
The suppression of this component in a bound nucleon is assumed to 
be the main source of the EMC effect in inclusive deep-inelastic
scattering \cite{FS88,FS85}. 
Note that this suppression does not lead to a noticeable change 
in the average characteristics of nucleons in nuclei \cite{FS85}.

To calculate the change of the probability of a PLC in a bound nucleon, 
one can use a perturbation series over a small parameter, $\kappa$, 
which controls corrections to the description of a nucleus as a system
of undeformed nucleons.
This parameter is taken to be the ratio of the characteristic energies 
for nucleons and nuclei: 
\begin{equation}
\kappa = \mid \langle U_A\rangle/\Delta E_A\mid \sim 1/10, 
\label{plc1}
\end{equation}
where $\langle U_A\rangle$ is the average potential energy per nucleon,
$\langle U_A\rangle\mid_{A\gg 1}\approx -40$~MeV, and 
$\Delta E_A \approx M^*~-~M \sim 0.5$~GeV is the typical energy 
for nucleon excitations within the nucleus. 
Note that $\Delta E_D \gtrsim 2(M_\Delta-M) \sim 0.6$~GeV, since 
the $N-\Delta$ component in the deuteron wave function is forbidden
due to the zero isospin of the deuteron.

To estimate the deformation of the bound nucleon wave function we 
consider a model where the interaction between nucleons is described 
by a Schr\"odinger equation with potential 
$V(R_{ij}, y_i, y_j)$ which depends both on the inter-nucleon distances 
(spin and isospin of nucleons) and the inner variables $y_i$ and $y_j$, 
where $y_i$ characterizes the quark-gluon configuration in the $i$-th 
nucleon \cite{FS88,FS85,FJM}.
The Schr\"odinger equation can be represented as:
\begin{equation}
\left[-{1\over 2m_N}\sum\limits_i\nabla^2_i + 
\sum\limits_{i,j}'V(R_{ij}, y_i, y_j) + \sum\limits_iH_0(y_i)\right]
\psi(y_i,R_{ij}) = E \psi(y_i, R_{ij}). 
\label{plc2}
\end{equation}
Here $H_0(y_i)$ is the Hamiltonian of a free nucleon. 
In the nonrelativistic theory of the nucleus the inter-nucleon 
interaction is averaged over all $y_i$. 
Thus the nonrelativistic $U(R_{ij})$ is related to $V$ as:
\begin{equation}
U(R_{ij}) = \sum\limits_{y_i,y_j,\widetilde y_i,\widetilde y_j} 
\langle \phi_N(y_i)\phi_N(y_j)\mid V(R_{ij},y_i,y_j,\widetilde y_i,
\widetilde y_j)\mid \phi_N(\widetilde y_i)\phi_N(\widetilde y_j)\rangle, 
\label{plc3}
\end{equation}
where $\phi_N(y_i)$ is the free nucleon wave function. 
The unperturbated wave function is the solution of Eq.(\ref{plc2}) 
with the potential $V$ replaced by $U$.  
Treating $(U-V) / (E_i - E_N)$ as a small parameter, where $E_i$ is
the energy of an intermediate excited nucleon state, one can calculate 
the dependence of the probability to find a nucleon in a PLC to the 
momentum of a nucleon inside the nucleus. 
One finds that this probability is suppressed as compared to the 
similar probability for a free nucleon by the factor \cite{FS85}:
\begin{equation}
\delta_A({\bf k}^2) \approx 1 - 4({\bf k}^2/2M + \epsilon_a)/\Delta E_A,
\label{plc6}
\end{equation}
where $\Delta E_A = \langle E_i - E_N \rangle \approx M^* - M$, 
in first order of the perturbation series. 
An estimate of higher order terms gives \cite{FS88}: 
\begin{equation}
\delta_A({\bf k}^2) = (1+z)^{-2}, \ \ \ \ \ 
z = ({\bf k}^2/M + 2\epsilon_A)/\Delta E_A. 
\label{plc7}
\end{equation}
The $x$ dependence of the suppression effect is based on the 
assumption that the PLC contribution in the nucleon wave function is 
negligible at $x \alt 0.3$, and gives the dominant contribution at 
$x \gtrsim 0.5$ \cite{FS85,FSS90}.  
We use a simple linear fit to describe the $x$ dependence between 
these two values of $x$ \cite{FSS90}.
One can then obtain an estimate for $R_A$ in Eq.(\ref{remc}) for large 
$A$ at $x\sim 0.5$, 
\begin{equation}
R_A(x)\mid_{x\sim 0.5} \sim \delta_A({\bf k}^2) 
\approx 1 + {4\langle U_A\rangle\over \Delta E_A} \sim 0.7-0.8 , 
\label{plct}
\end{equation}
since here Fermi motion effects are small.
The excitation energy $\Delta E_A$ for the compressed configuration 
is estimated as 
$\Delta E_A \sim (M(1400)-M) - (M(1680)-M)
            \sim$ (0.5--0.8)~GeV, 
while $\langle U_A\rangle\approx -40$~MeV.
Since $\langle U_A\rangle\sim \langle \rho_A(r)\rangle$ the model 
predicts also the $A$ dependence of the EMC effect, which is 
consistent with the data \cite{FS88}.

For the deuteron target we can deduce from Eq.(\ref{plct}) using   
$\langle U_D\rangle/\langle U_{Fe}\rangle\sim 1/5$,
\begin{equation}
R_D(x,Q^2)\mid_{x\sim 0.5}\approx 0.94 - 0.96.
\label{emcd}
\end{equation}
This number may be somewhat overestimated because, as discussed above,
due to the isoscalarity of $D$, low-energy excitations in the 
two-nucleon system are forbidden, leading to $\Delta E_D > \Delta E_A$.

This model represents one of the extreme possibilities that the EMC 
effect is solely the result of deformation of the wave function of 
bound nucleons, without attributing any extra momentum to be carried
by mesons.
A distinctive feature of this explanation of the EMC effect is that 
the deformation of the nucleon should vary with inter-nucleon distances
in nuclei (with nucleon momentum in the nucleus). 
For $|{\bf k}| \sim 0.3-0.4$ GeV/c the deviation from the conventional 
quantum-mechanical model of a deuteron is expected to be quite large 
(factor $\sim 2$). 
Actually, the size of the effects may depend not on ${\bf k}^2$ only 
but on $p_T$ and $\alpha_s$ separately, because the deformation of a 
bound nucleon may be more complicated than suggested by this simple model.

\subsection{QCD Radiation, Quark Delocalization.}

It was observed in Refs.\cite{CLRR,NP} that the original EMC data could be 
roughly fitted as:
\begin{equation}
{1\over A} F_{2A}(x,Q^2) = {1\over 2} F_{2D}\left(x,Q^2\xi_{A}(Q^2)\right), 
\label{eq.5.39}
\end{equation}
with $\xi_{Fe}(Q^2) \approx 2$ for $Q^2\approx 20$~GeV$^2$, the so-called 
dynamical rescaling. 
The phenomenological observation has been interpreted as an indication that 
gluon radiation occurs more efficiently in a nucleus than in a free nucleon 
(at the same $Q^2$) due to quark delocalization, either in a bound nucleon 
(or in two nearby nucleons) \cite{CLRR,JRR,CLOSE,GP} or in the nucleus as a 
whole \cite{JRR}.

The $Q^2$ dependence of $\xi(Q^2)$ follows from the requirement that both
sides of Eq.(\ref{eq.5.39}) should satisfy the QCD evolution equations.
In the leading logarithmic approximation one has:
\begin{equation}
\xi_A(Q^2) = \xi_A(Q_0^2)^{\alpha_{QCD}(Q_0^2)/\alpha_{QCD}(Q^2)}.
\label{eq.5.40}
\end{equation}
Experimentally $d \ln F_{2D}(x,Q^2)/d \ln Q^2$ is positive if $x>x_0$ and
negative if $x<x_0$, where $x_0 = 0.15 \pm 0.05$. 
So Eq.(\ref{eq.5.39}) predicts that the EMC effect should vanish 
at $x\approx x_0$.

If the confinement size in a nucleus, $\lambda_A$, is larger than that
in a free nucleon, one may expect that the $Q^2$ evolution of the parton
distributions in nuclei (the bremsstrahlung of gluons and quarks) may
start at $Q_0^2(A) < Q_0^2(N)$. 
To reproduce Eq.(\ref{eq.5.39}) one should have:
\begin{equation}
Q_0^2(A)/Q_0^2(N) = [\xi(Q_0^2(A))]^{-1}.
\label{eq.5.41}
\end{equation}
Assuming on dimensional grounds that the radii of quark localization in
a nucleus, $\lambda_A$, and in a nucleon, $\lambda_N$, are related via:
\begin{equation}
Q_0^2(A)\lambda_A^2 = Q_0^2(N)\lambda_N^2
\label{eq.5.42}
\end{equation}
to the scale for the onset of evolution, $Q_0^2$,
one finally obtains:
\begin{equation}
\lambda_A/\lambda_N \approx {\xi(Q_0^2(A))}^{1/2}. 
\label{eq.5.43}
\end{equation}
A fit to the original EMC data using Eq.(\ref{eq.5.39}) leads to 
$\xi_{Fe}$(20 GeV$^2$)=2. 
For $\mu_{Fe}^2 = 0.67$~GeV$^2$ and
$\Lambda_{\overline{MS}} \approx 250$~MeV this corresponds to: 
\begin{equation}
\lambda_{Fe}/\lambda_D \approx 1.15 \ \ \ \ \ \ \ \ 
\lambda_{A\sim 200}/\lambda_D \approx 1.19. 
\label{eq.5.44}
\end{equation}

Since in this model the delocalization is approximately proportional to the 
nuclear density, one can expect that the effect is also proportional to the 
${\bf k}^2$ of a bound nucleon. 
Fixing the parameters of the model to fit the Fe data, and assuming 
$\lambda(k)/\lambda = 1 + a {\bf k}^2$ we obtain 
$a\approx 0.4/<{\bf k}^2>_{Fe}$,  
where $<{\bf k}^2>_{Fe} \sim 0.08$~GeV$^2$/c$^2$.
Using this expression for $\lambda(k)$ one can calculate the $\xi_d(Q^2,k)$ 
according to Eqs.(\ref{eq.5.40})--(\ref{eq.5.43}) and express the effective 
structure function of a bound nucleon as:
\begin{equation}
F_{2N}^{eff}(x,\alpha_s,p_T,Q^2) = F_{2N}\left(x,Q^2\xi(Q^2,k)\right), 
\label{rsc}
\end{equation}
where $k$ is defined through $\alpha_s$ and $p_T$ according to Eq.(\ref{k}).

\subsection{Numerical estimates}
\label{numcom}

Comparison of predictions of the models for the nuclear EMC effect 
considered in the preceding Sections is most meaningful in the kinematic
range where, firstly, FSI effects are small, and secondly, the instant-form 
and light-cone prescriptions for the deuteron spectral function lead to
similar results.

Direct calculation of the FSI contribution to the cross section would 
require knowledge of the full dynamics of the final $N$--$X$ system,
which is a practically impossible task given the present level of
understanding of nonperturbative QCD. 
However, it is possible to estimate the uncertainty which would be 
introduced through neglect of FSI, by using the calculation of FSI effects
in the high-energy $d(e,e'p)n$ (break-up) reaction in
Refs.\cite{FMGSS95,FMGSS96}, and replacing the $p$--$n$ rescattering 
cross section by an effective cross section for the $p(n)$--$X$
interaction, Eq.(\ref{e7}). 
For the effective cross section $\sigma_{eff}$ one can use the results of
the recent analysis \cite{STZ} of soft neutron production in the high-energy
deep-inelastic scattering of muons from heavy nuclei \cite{E665}, which
yielded an upper limit of $\sigma_{eff} \approx 20$ mb.
To be on the conservative side, in the following estimates we therefore
use the value of $\sigma_{eff} \approx 20$ mb suggested by string models
of FSI (for a recent summary see Ref.\cite{BORIS}).
Furthermore, by retaining only the imaginary part of the spectator nucleon
rescattering amplitude, one obtains an upper limit of the FSI effect,
since the real part will contribute to elastic rescattering only,
effectively suppressing the value of $\sigma_{eff}$.
In Fig.1 we illustrate the $\alpha_s$ dependence of the ratio of the
(light-cone) spectral function including FSI effects within the DWIA,
Eq.(\ref{e7}), to that calculated without FSI effects.
At extreme backward kinematics $(p_T \approx 0)$ one sees that FSI
effects contribute less than $\sim 5\%$ to the overall uncertainty
of the $d(e,e'N)X$ cross section for $\alpha_s \alt 1.5$.
As mentioned above, this number can be considered rather as an upper
limit on the uncertainties due to FSI.
At larger $p_T$ ($\agt 0.3$ GeV/c), and small $\alpha_s$ ($\approx 1$),
the double scattering contribution (which is not present for the
extreme backward case in Eq.(\ref{e7})) plays a more important role
in FSI \cite{FMGSS95}.
Because its sign is positive, it tends to cancel some of the absorption
effects of FSI at large $p_T$ (for a detailed discussion of the double
scattering contribution in FSI see Ref.\cite{FMGSS95}).
Note also that the FSI effects do not change significantly with energy
for fixed $Q^2$.
Thus, for the ratios discussed, where the cross sections are compared
for the same $Q^2$ but different $x$, the changes due to FSI effects
are even smaller.

To extract unambiguous information from the semi-inclusive cross section
ratios on the medium modifications of the nucleon structure discussed
in this Section requires one to establish the regions of kinematics where
the differences between the various prescriptions for the deuteron spectral
function are minimal.
In Fig.2 we illustrate the $\alpha_s$ and $p_T$ dependence of the ratio of 
spectral functions calculated in the light-cone (\ref{lcs}) and instant-form 
(\ref{Sif}) approaches.
For $p_T \le 0.1$~GeV/c the light-cone and instant-form predictions differ 
up to the 20\% for the entire range of $\alpha_s \le 1.5$. 
However, choosing the isolated values of $\alpha_s \le 1.2$ or 
$\alpha_s \approx 1.4$ one can confine the uncertainty in the spectral 
function to within 10\%, which will offer the optimal conditions in which
to study the nucleon structure modification.

In Fig.3 we show the $\alpha_s$ dependence of the ratio of the effective 
proton structure function, $F^{eff}_{2p}$, for extreme backward kinematics,
$p_T=0$, to the structure function of a free proton. 
As expected, the suppression of the ratio in the version of the binding 
model with explicit nucleon off-shell corrections is quite small, 
$\alt 10\%$ for $\alpha_s < 1.5$, reflecting the relatively minor role 
played here by off-shell effects in the nucleon structure function.
(Note that in versions of the binding model in which there is no $p^2$ 
dependence in the effective nucleon structure function this ratio would 
be unity.) 
The effects in the PLC suppression and rescaling models in Fig.3 are 
somewhat larger.
For a neutron target one predicts similar results, however, the neutron
structure function is not as well known experimentally as the proton due
to the absence of free neutron targets (see Section IV).
For this reason we restrict our discussion to ratios of proton structure
functions only.

In Fig.4 we illustrate the dependence of $F^{eff}_{2p}/F_{2p}$ on the
variable $(p^2 - 2M\epsilon)/M^2$, where $p^2$ is the bound nucleon 
virtuality defined in Eq.(\ref{virt}) and $\epsilon = -2.2$ MeV is the
deuteron binding energy.
The comparison in Fig.4 is done for different values of $x$ and $\alpha_s$. 
It is noticeable that because of the negligible amount of the PLC component
in the nucleon wave function at $x\lesssim 0.3$, the PLC suppression model
predicts no modification of the structure function in this region. 
On the other hand, at $x=0.6$ it predicts a maximal effect because of 
the PLC dominance in the nucleon wave function here.

Because the momentum (virtuality) dependent density effect generates 
the modification of the bound nucleon structure functions in the PLC 
suppression and rescaling models, the ratios for these models at 
$x\gtrsim$ 0.5--0.6 vary similarly with $\alpha_s$ and 
$(p^2 - 2M\epsilon)/M^2$, as does also the off-shell model ratio.
However the mechanism for such a variation is different, which is clearly 
seen in Fig.5, where the $x$ dependence of the same ratio is represented
at different values of $\alpha$ and fixed $p_T=0$.
The curves for the off-shell model extend only to $x \sim 0.7$ because
for larger $x$ values the approximations discussed in Sec.\ref{IF} 
involved in obtaining Eq.(\ref{ia}) become numerically less justified 
\cite{MST,MSTD}.

To further reduce any uncertainties due to the deuteron spectral function
in the model comparisons, we concentrate on the predictions for the ratio 
$G(\alpha_s,p_T,x_1,x_2,Q^2)$, defined in Eq.(\ref{G}), of experimentally
measured cross sections at two different values of $x$ \cite{FS85}.
Since the function $G$ is defined by the ratio of cross sections at the 
same $\alpha_s$ and $p_T$, any uncertainties in the spectral function 
cancel.
This allows one to extend this ratio to larger values of $\alpha_s$,
thereby increasing the utility of the semi-inclusive reactions when 
analyzed in terms of this function.
Figure 6 shows the $\alpha_s$ and $Q^2$ dependence of
$G(\alpha_s,p_T,x_1,x_2,Q^2)$ at $p_T=0$.
The values of $x_1$ and $x_2$ are selected to fulfill the condition 
$x_1/(2-\alpha_s)=0.6$ (large EMC effect in inclusive measurements) 
and $x_2/(2-\alpha_s)=0.2$ (essentially no EMC effect in inclusive 
measurements). 
Again, the PLC suppression and $Q^2$ rescaling models predict a much 
faster drop with $\alpha_s$ than does the binding/off-shell model, 
where the $\alpha_s$ dependence is quite weak.

\section{Extraction of the Neutron/Proton Structure Function Ratio}

The presence of an EMC effect in the deuteron leads to substantial 
suppression of the deuteron structure function at large $x$ compared 
to what one would expect from models without non-nucleonic degrees 
of freedom. 
For example, the estimate of Ref.\cite{FS85}, 
\begin{equation}
\left.
{F_{2D}(x,Q^2) \over F_{2p}(x,Q^2)+F_{2n}(x,Q^2)} 
\approx {1 \over 4} {2 F_{2A}(x,Q^2)\over A F_{2D}(x,Q^2)}
\right|_{A \sim 60; 0.3 < x < 0.7}
\end{equation}
which is valid for a rather wide class of models in which the EMC effect 
is proportional to the mean value of $p^2$ in nuclei, gives a ratio 
$\sim$ 3--5\% below unity at $x \sim$ 0.6--0.7.
Calculations of the deuteron structure function in models in which binding 
effects are explicitly taken into account also produce similar effects 
\cite{MSTD}.

Inclusion of the EMC effect in the extraction of the the neutron structure
function from the inclusive $e D$ scattering data leads to significantly
larger values for $F_{2n}/F_{2p}$ than the 1/4 value obtained in analyses 
in which only Fermi motion is included.
The values for $F_{2n}/F_{2p}$ with inclusion of the EMC effect at 
$x \sim 0.6$ \cite{MT,BODEK,STR92} are in fact much closer to the 
expectation of 3/7 from perturbative QCD, predicted by Farrar and Jackson 
\cite{FJ}.
Therefore observation of a value of $F_{2n}/F_{2p}$ higher than the 1/4 
extracted from inclusive data in the early analyses would by itself serve
as another proof of the presence of the EMC effect in the deuteron.

Although other methods to obtain the large-$x$ $n/p$ ratio (or the $d/u$ 
ratio) have been suggested, none has so far been able to clearly discriminate
between the different $x \rightarrow 1$ limits for $F_{2n}/F_{2p}$ (namely,
$d/u \rightarrow 0$, which is the minimal possible value allowed in the
parton model, which corresponds to $F_{2n}/F_{2p} \rightarrow 1/4$, and 
$d/u \rightarrow 1/5$ in perturbative QCD \cite{FJ}).
For example, with $\nu$ and $\overline\nu$ beams on proton targets one can 
in principle measure the $u$ and $d$ quark distributions separately, however,
the statistics in $\nu$ experiments in general are relatively poor.
Another possibility would be to extract the $d/u$ ratio from charged lepton 
asymmetries at large rapidities, in $W$-boson production in $p\overline p$ 
scattering \cite{W}, although here it may be some time still before a 
sufficient quantity of large-rapidity events at the CDF at Fermilab are
accumulated.

On the other hand, with tagged deuteron experiments planned for HERMES, 
a study of the tagged structure functions may allow a resolution of this 
ambiguity \cite{KG,FS81,SIM}.
It is important that the ratio of tagged structure functions interpolated
to the nucleon pole should be exactly equal to the free nucleon ratio --- 
this is the analog of the Chew-Low interpolation for the pion case:  
\begin{eqnarray}
\label{F2np}
{ F_{2n}(x,Q^2) \over F_{2p}(x,Q^2) }
&\approx& 
\left. 
{ F_{2n}^{eff}(x/(2-\alpha_s),p_T,Q^2) \over 
  F_{2p}^{eff}(x/(2-\alpha_s),p_T,Q^2) }
\right|_{\alpha_s \approx 1, p_T \approx 0}.  
\end{eqnarray}
In practice the data cannot be accumulated for too small ${\bf p}$.
However, we observed above that deviations of the ratio from the free 
limit is proportional to $p^2$ with a good accuracy.
Hence, if one samples the data as a function of $p^2$ interpolation 
to the pole $p^2-M^2=2M\epsilon$ should be smooth. 
In practice, considering momentum intervals of 100--200 MeV/c and 
200--350 MeV/c would be sufficient. 
A potential problem with Eq.(\ref{F2np}) is that at very large $x$
($x \agt 0.7$) the factorization approximation itself breaks down and 
higher order corrections to Eq.(\ref{ia}), which are $\propto {\bf p}^4$, 
must be included if one wants accuracy to within a few \%.
To be on the safe side one should therefore restrict the analysis 
to smaller spectator momenta, below 200 MeV/c.

\section{Conclusion}

Despite the many years of study of the deviations from unity of the 
ratios of nuclear to deuteron cross sections in inclusive high-energy
scattering, we have been unable to isolate the microscopic origin of
the nucleon structure modification in the nuclear environment 
\cite{EMC1,EMC2,EMC3,EMC4,EMC5}
--- the effect can be described in terms of a number of models based 
on quite disparate physical assumptions.
In this paper we have argued that by allowing greater accessibility 
to kinematic variables not available in inclusive reactions, 
semi-inclusive deep-inelastic processes offer the possibility to 
make further progress in understanding the origin of the nuclear
EMC effect.

In particular, measurements of tagged structure functions of the 
deuteron can probe the extent of the deformation of the intrinsic 
structure of a bound nucleon. 
Taking ratios of semi-inclusive cross sections at different values of 
$x$ enables the cancellation in the tagged structure function ratio,
Eq.(\ref{G}), of the dependence on the deuteron wave function, thus 
permitting the nucleon structure to be probed directly.  
Our results show that possible contamination of the signal due to the
final state interactions of the spectator nucleon with the hadronic
debris can be minimized by tagging only on the slow backward nucleons
in the target fragmentation region.
This may then allow one to discriminate between models in which the 
EMC effect is attributed entirely to the nucleon wave function 
deformation, and ones in which the effect arises from more traditional 
descriptions in terms of meson--nucleon degrees of freedom associated 
with nuclear binding.

As a by-product of the semi-inclusive measurements, one may also be 
able to extract information on the large-$x$ behavior of the neutron 
to proton structure function ratio, by detecting recoiling protons 
and neutrons with small transverse momentum in the extreme backward 
kinematics.
Extracting this ratio from inclusive $e p$ and $e D$ data is fraught
with large uncertainties arising from different treatments of the 
nuclear physics in the deuteron.
Observation of an asymptotic value for $F_{2n}/F_{2p}$ which is larger 
than the `canonical' 1/4 would in itself be proof of the presence of 
an EMC effect in the deuteron.

First information about the spectra of backward nucleons in $eD$ scattering 
is likely to come from the Jefferson Lab experiment \# 94-102 \cite{KG}, 
and from the HERMES experiment at HERA, where the necessary counting rate
may be achieved after an upgrade of the detector \cite{VS}. 
Having two energy ranges would be very useful for checking the basic 
production mechanism, and understanding backgrounds and corrections due
to final state interactions, which are likely to depend substantially on 
the incident energy.
Final state interaction effects can also be tracked by studying the 
production of nucleons from heavier nuclei.
Probing the quark-gluon structure of short-range correlations with 
heavy nuclei targets should further enable one to determine whether 
nucleon deformations predominantly depend on the nucleon momentum, 
or also on $A$.

\acknowledgements

We would like to thank L.L. Frankfurt, A.W. Thomas and G. van der Steenhoven
for many useful discussions and suggestions.
We thank the Institute for Nuclear Theory at the University of Washington 
for its hospitality and support during recent visits, where part of this 
work was performed.
This work was supported by the U.S. Department of Energy grants
DE-FG02-93ER-40762, DE-FG02-93ER-40771 and by the U.S.A. -- Israel 
Binational Science Foundation Grant No. 9200126.


\begin{figure}
\caption{The $\alpha_s$ dependence of the ratio of cross sections
	calculated with FSI effects within the DWIA, and without FSI effects. 
	The curves correspond to different values of the spectator
	nucleon transverse momentum (in GeV/c).}
\end{figure}

\begin{figure}
\caption{The $\alpha_s$ dependence of the ratio of cross sections
	calculated within the light-cone and instant-form approaches.
	The curves correspond to different values of the 
	spectator nucleon transverse momentum (in GeV/c).}
\end{figure}

\begin{figure}
\caption{The $\alpha_s$ dependence of $F_{2p}^{eff}/F_{2p}$
	for $x=0.6$ and $p_T=0$.
	Dashed line is the PLC suppression model, dotted is the
	rescaling model, and dot-dashed the binding/off-shell model.}
\end{figure}

\begin{figure}
\caption{The $(p^2-2M\epsilon)/M^2$ dependence of $F_{2p}^{eff}/F_{2p}$,
	for $\alpha_s = 1.2$ and 1.4, and $x=0.3$ and 0.6.
	Curves are as in Fig.3.}
\end{figure}

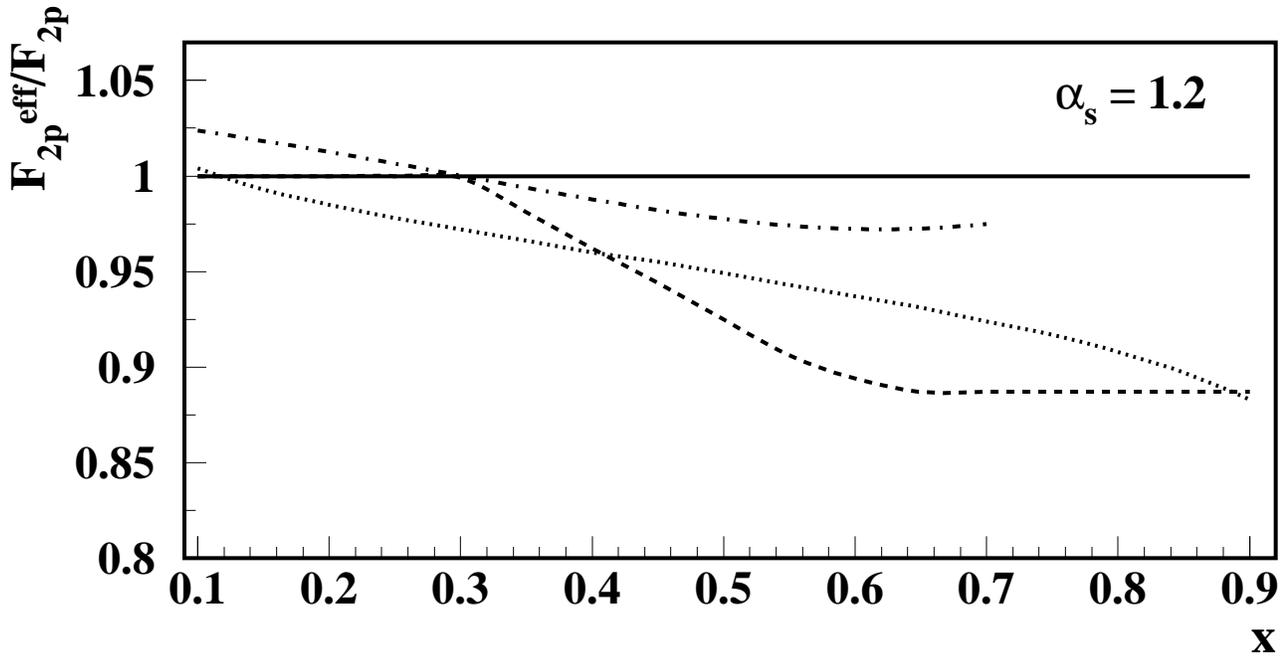
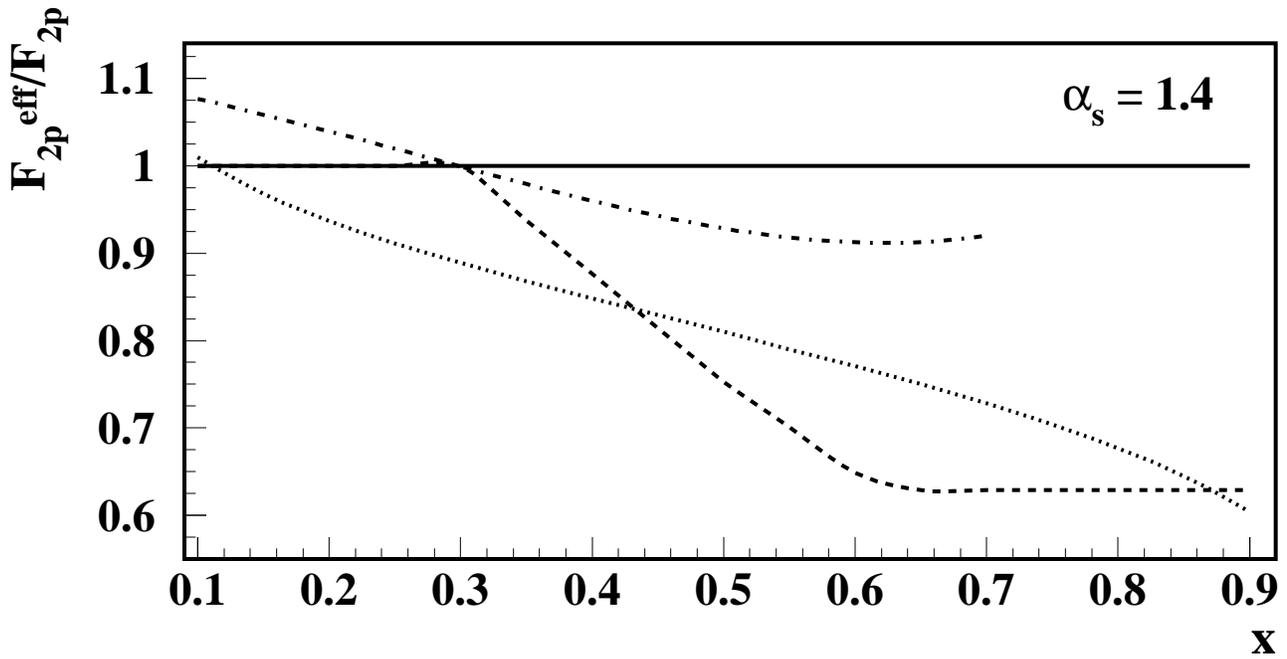
\begin{figure}
\caption{The $x$ dependence of $F_{2p}^{eff}/F_{2p}$ for 
	$\alpha_s = 1.2$ and 1.4, with $p_T=0$.
	Curves are as in Fig.3.}
\end{figure}

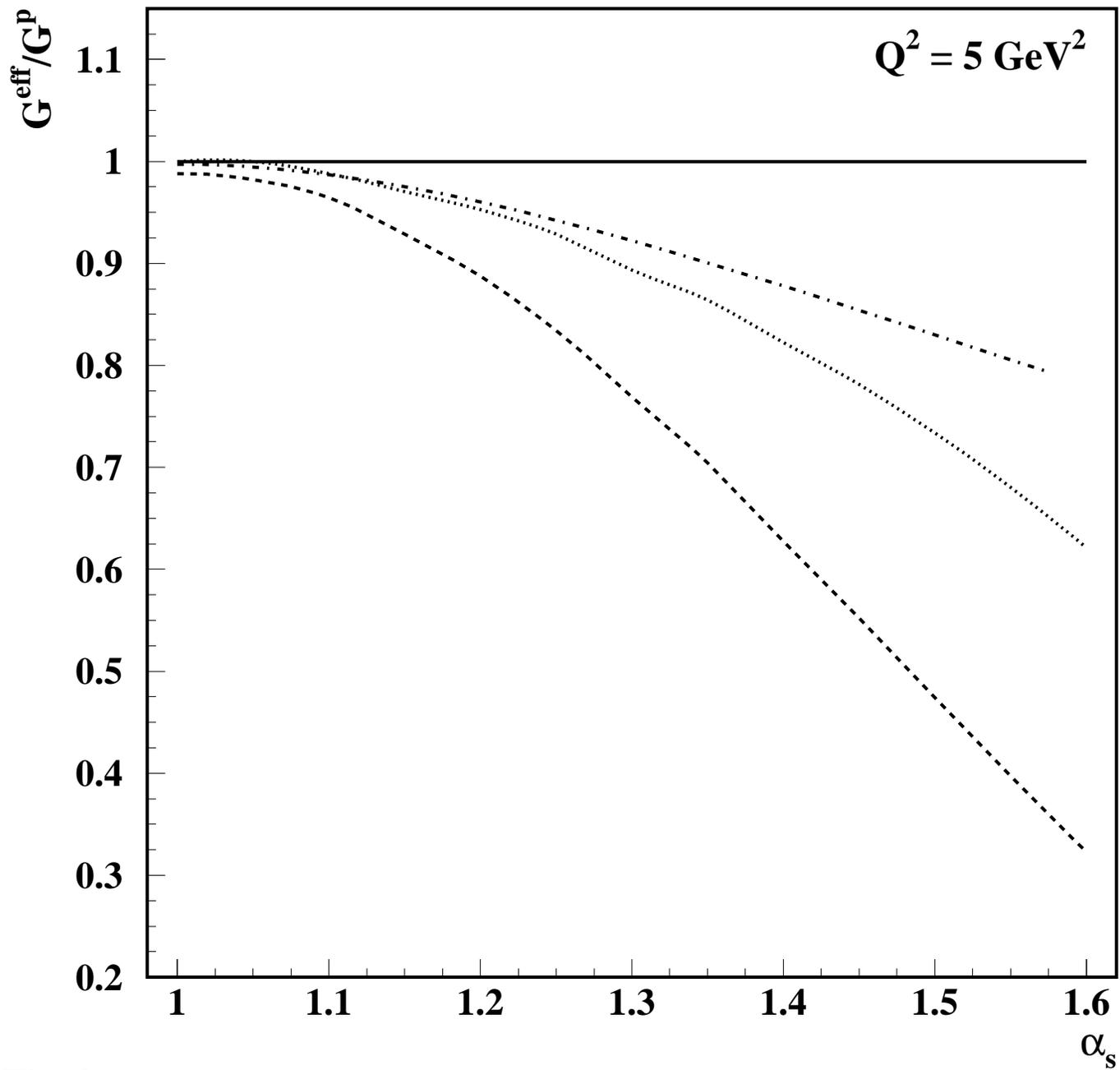
\begin{figure}
\caption{The $\alpha_s$ dependence of $G(\alpha_s,p_T,x_1,x_2,Q^2)$, 
	with $x_1=x/(2-\alpha_s)=0.6$ and $x_2=x/(2-\alpha_s)=0.2$,
	for $p_T=0$.
	$G^{eff}(\alpha_s,p_T,x_1,x_2,Q^2)$ is normalized to 
	$G^{eff}(\alpha_s,p_T,x_1,x_2,Q^2)$
	calculated with the free nucleon structure function. 
	Curves are as in Fig.3.}
\end{figure}

\end{document}